\documentclass[aps,prl,twocolumn,groupedaddress,floatfix]{revtex4}

\usepackage{graphics}

\def\lb{\left(}
\def\rb{\right)}
\def\be {\begin{equation}}
\def\ee {\end{equation}  }
\def\beq{\begin{eqnarray}}
\def\eeq{\end{eqnarray}  }
\def\bi {\begin{itemize} }

\def\ei {\end{itemize}   }
\def\RE {I\kern-6pt R    }
\def\Z  {Z\kern-13pt Z   }
\def\be {\begin{equation}}
\def\ee {\end{equation}  }
\def\beq{\begin{eqnarray}}
\def\eeq{\end{eqnarray}  }

\def\eeq{\end{eqnarray}  }

\begin{document}

\title{The Singularity Threshold of the Nonlinear Sigma Model Using 3D Adaptive Mesh Refinement}

\author{Steven L. Liebling}
    %\email[]{steve@mozart.liu.edu}
    %\homepage[]{http://techcenter.southampton.liu.edu/steve}
    %\thanks{}
    %\altaffiliation{}
    \affiliation{Theoretical and Computational Studies Group\\
                 Southampton College, Long Island University\\
                 Southampton, NY 11968}

\date{\today}

\begin{abstract}
Numerical solutions to the nonlinear sigma model (NLSM), a wave map from
$3+1$ Minkowski space to $S^3$, are computed in three
spatial dimensions (3D) using adaptive mesh refinement (AMR).
For initial data with compact support the model
is known to have two regimes, one in which regular initial data forms
a singularity and another in which the energy is dispersed to infinity.
The transition between these regimes has been shown in
spherical symmetry to demonstrate threshold behavior
similar to that between black hole formation and dispersal in gravitating
theories. Here, I generalize the result by removing the assumption of
spherical symmetry. The evolutions suggest that the spherically symmetric
critical solution remains an intermediate attractor
separating the two end states.
\end{abstract}


\maketitle

%%%%%%%%%%%%%%%%%%%%%%%%%%%%%%%%%%%%%%%%%%%%%%%%%%%%%%%%%%%%%%%%
%{\em Introduction:}
%%%%%%%%%%%%%%%%%%%%%%%%%%%%%%%%%%%%%%%%%%%%%%%%%%%%%%%%%%%%%%%%
Nonlinear sigma models have attracted the attention both of physicists
and mathematicians. For physicists, the models represent the possibility
of describing particles with a field theory, while mathematicians
study singularity formation in geometrically motived nonlinear
models.
Recently, a nonlinear sigma model has attracted the
attention of relativists because it demonstrates behavior
similar to black hole critical phenomena discovered by
Choptuik~\cite{Choptuik:1993jv}.

Studying the gravitational collapse of a spherically symmetric scalar
field, he found that initial data with large energy formed a black
hole while that for small initial energy dispersed its energy to
infinity. By tuning the initial energy, he approached the so-called
critical regime which separates these two end states. In this regime,
solutions approach a unique, universal solution which demonstrates
self-similarity.

The model studied here serves as a useful toy model
for such threshold behavior because it also possesses two stable
end-states, dispersal and singularity formation. These end-states
serve as analogues for the gravitating case in which initial data
can either disperse or form a black hole.
Similar to Choptuik's gravitating model,
spherically symmetric studies of this model have found
a universal, self-similar critical solution~\cite{Liebling:1999nn,bizon:2000}.
The utility of the nonlinear sigma model arises from 
its relative simplicity compared to any of a variety of gravitating
models. Perhaps the study of such nonlinear, flat
systems will guide the way to further insight in the gravitating models.

While questions about the nature of black hole critical behavior
remain, many unexplored ``experiments'' have yet
to be conducted.
While many matter models have been studied (for
a review see~\cite{Gundlach:1999cu}), generally these studies have been
restricted to spherical symmetry. The only work to model 
fully nonlinear collapse in less symmetry
is that of Abrahams and Evans~\cite{Abrahams:1993wa} who studied vacuum
gravitational collapse in axisymmetry.
One reason for the scarcity of multi-dimensional evolutions
of the critical regime is 
the extreme resolution demanded by approach to
scale invariance.
Perturbation methods are a complementary approach
with recent work~\cite{Gundlach:1999cw,Martin-Garcia:1998sk} suggesting that
spherical critical solutions in the scalar field and perfect fluid
cases extend to the nonspherical regime.
While work is underway to duplicate and extend the work
of Abrahams and Evans in axisymmetry~\cite{us}, it makes sense
to look ahead to a simple model in three spatial dimensions.

To obtain the resolution required for evolving self-similar
solutions, Choptuik developed a computational infrastructure in
one spatial dimension which
dynamically and locally adds numerical resolution where needed using
adaptive mesh refinement~(AMR). With this infrastructure, fine
subgrids are added and subtracted to the computational domain
providing resolution only where and when needed.
Because the computational cost scales as a power law in the spatial
dimension ({\em i.e.} doubling the resolution of a $d$-dimensional
evolution requires  a factor $2^{d+1}$ more work),
it is expected that
AMR will be absolutely crucial in higher dimensions
for well-resolved evolutions
of interest, in particular
black hole critical phenomena or black hole collisions.

Constructing such a gravitating model with AMR is an ambitious
project. Instead I report on the construction of a
3D AMR code of the nonlinear sigma model,
which genuinely requires AMR 
and holds physical interest.

%%%%%%%%%%%%%%%%%%%%%%%%%%%%%%%%%%%%%%%%%%%%%%%%%%%%%%%%%%%%%%%%
{\bf \em The NLSM Model:}
%%%%%%%%%%%%%%%%%%%%%%%%%%%%%%%%%%%%%%%%%%%%%%%%%%%%%%%%%%%%%%%%
The nonlinear sigma model studied here represents a mapping from
the base space of $3+1$ Minkowski
to a target space of $S^3$. In spherical symmetry, it is common to
choose the hedgehog ansatz for the map reducing the dynamics to that
of a single spherically symmetric field $\chi(r,t)$.
Here a simple generalization of this ansatz is chosen
\be
\phi^a = \left( \begin{array}{c}
                \sin \chi(x,y,z,t) \sin \theta \sin \varphi \\
                \sin \chi(x,y,z,t) \sin \theta \cos \varphi \\
                \sin \chi(x,y,z,t) \cos \theta \\
                \cos \chi(x,y,z,t)
                \end{array} \right),
\label{eq:phi}
\ee
where $\theta$ and $\varphi$ are the usual spatial
angles.  The dynamics
reduce to the scalar field $\chi(x,y,z,t)$
which satisfies the equation of
motion
\be
\ddot \chi =
                   \chi_{,xx}
                  + \chi_{,yy}
                  + \chi_{,zz}
             -\frac{ \sin 2\chi}{r^2},
\label{eq:eom}
\ee
where commas indicate partial differentiation with respect to subscripted
coordinates, an overdot denotes $\partial / \partial t$, and $r \equiv
\sqrt{x^2+y^2+z^2}$. The equation of motion~(\ref{eq:eom}) implies the
regularity condition $\chi(0,0,0,t)= 0$ which is enforced by the evolution
procedure. This ansatz requires that the origin is singled out as a special
point.  More elegant generalizations
may be considered in the future.

The energy density of the map is given by
\be
\rho = \frac{1}{2} \left[
                                     \lb \dot \chi \rb^2
                                   + \lb \chi_{,x} \rb^2
                                   + \lb \chi_{,y} \rb^2
                                   + \lb \chi_{,z} \rb^2
                                   \right]
                + \frac{\sin^2 \chi}{r^2}.
\ee
The angular momentum densities are given by~\cite{ryder}
\be
M^{\mu \nu} = \int d^3 x \left( T^{0\mu} x^\nu - T^{0\nu} x^\mu \right),
\ee
so that the $z$-component of the angular momentum, for example, is
\be
J_z = \int d^3 x~M^{xy} = \int d^3 x ~\dot \chi \left( y \chi_{,x} - x \chi_{,y} \right).
\ee
While the map allows for the possibility of a texture charge 
associated with the third homotopy group, only initial data with
zero charge is considered here. The model requires
initial data $\chi(x,y,z,0)$ and $\dot \chi(x,y,z,0)$
be specified at the initial time.

Various types of initial data have been implemented
and are described in
Table~\ref{table:init}. Some of these families are defined
in terms of a generalized Gaussian pulse defined by
\be
     G(x,y,z)  =  A e^{-(\tilde r - R)^2 / \delta^2},
\label{eq:id}\\
\ee
where $\tilde r$ is a generalized radial coordinate
\be
\tilde r = \sqrt{ \epsilon_x \left(x-x_c\right)^2
                + \epsilon_y \left(y-y_c\right)^2
                +            \left(z-z_c\right)^2 }.
\ee
Such a pulse depends on parameters: amplitude $A$, shell radius $R$,
pulse width $\delta$, pulse center $\left(x_c,y_c,z_c\right)$,
 and skewing factors $\epsilon_x$ and $\epsilon_y$.
For $\epsilon_x \ne 1 \ne \epsilon_y$ such a pulse has elliptic
cross section.
Family (a) represents a single
pulse for which 
the parameter $\nu$ takes the values
$\{-1,0,+1\}$ for an approximately out-going, time-symmetric, or approximately
in-going pulse.
The angular momentum of the pulse about the $z$-axis
is proportional to the parameter $\Omega_z$ as well as to $\left(\epsilon_x - \epsilon_y\right)^2$.

\begin{table}[h]
     \begin{tabular}{ | l | l | c | c |}
     \hline
     & Description & $\chi(x,y,z,0)$ & $\dot \chi(x,y,z,0)$ \\
     \hline
     \hline
     a & Ellipsoid            & $ G$
                          & $   \nu      \frac{\partial G}{\partial \tilde r}$\\
       &                  & ~ & $ + \Omega_z \left(y G_{,x} - x G_{,y} \right)$ \\
     \hline
     b & Two pulses   & $ G_1 + G_2 $
                          & $ v_1 \frac{\partial G_1}{\partial x} + v_2 \frac{\partial G_2}{\partial x}$ \\
     \hline
     c & Toroid   & $ A e^{-z^2/\delta^2}e^{ -\left(\epsilon_x x^2 + \epsilon_y y^2\right)^2 / \delta^2 }$
                          & $0$ \\
     \hline
     \end{tabular}
\caption{\label{table:init}List of various initial data families. 
         For families (a)-(c) both the field $\chi(x,y,z,0)$ and its
         time derivative $\dot \chi(x,y,z,0)$ are shown in terms of various
         parameters. The terms $G$, $G_1$, and $G_2$ represent unique Gaussian pulses
         as defined in Eq.~(\ref{eq:id}). In family (b), the parameters
         $v_1$ and $v_2$ are the respective velocities of the two pulses, generally
         chosen to have a grazing collision.}
\end{table}

%%%%%%%%%%%%%%%%%%%%%%%%%%%%%%%%%%%%%%%%%%%%%%%%%%%%%%%%%%%%%%%%
{\bf \em Numerical method:}
%%%%%%%%%%%%%%%%%%%%%%%%%%%%%%%%%%%%%%%%%%%%%%%%%%%%%%%%%%%%%%%%
RNPL~\cite{rnpl} is used to develop and debug a stable and convergent
unigrid code which solves Eq.~(\ref{eq:eom})
using finite differences and iterative Crank-Nicholson.
The RNPL generated update procedure is 
called from the AMR code to evolve any given fine grid.
The AMR implementation follows that of Berger and Oliger~\cite{berger}
with some simplifications:

(1)~Instead of estimating truncation error, I use the energy density
      as a criterion of refinement. In particular, normalizing the
      energy density $\rho$ by the grid resolution $1/h$, the refinement
      criterion is $h^2 \rho > \epsilon$, where $\epsilon$ is
      a user-specified threshold.
      This simplification is less general than truncation error but
      provides an easily computable and smooth function from which
      to estimate resolution requirements.

(2)~Fine grids are completely contained
      within their parents. Grids do not
      overlap or abut other grids at the same level.
      This restriction reduces the maximum obtainable efficiency but provides for
      considerable simplification.

(3)~Fine grids are created strictly aligned with parent grids with
      no rotation. The transformations necessary to relax this would
      be especially onerous in the gravitating case.

(4)~The ratio of refinement between parent and child grids is
      constrained to be an even integer.

The first two of these restrictions appear especially suited for
the case of central collapse studied here, and
the intention is to relax them for more general problems.

%%%%%%%%%%%%%%%%% Figure:
\begin{figure}
\vspace{-10mm}
\centerline{\scalebox{0.45}{\includegraphics{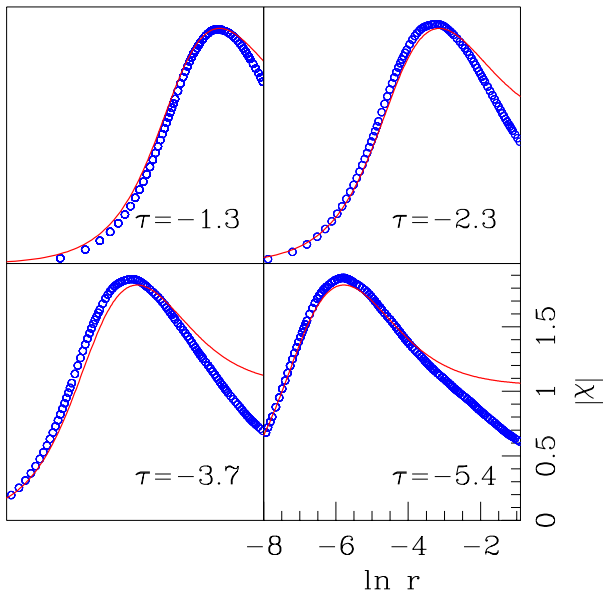}}}
\vspace{-3mm}
\caption{\label{fig:ss}Approach of near-critical evolution
    to self similarity. The numerical evolution is
    shown~(circles) for spherically symmetric initial data
    at times near the collapse time. The initial data is a member of
    family (a) in Table~\ref{table:init}
    with $\epsilon_x = \epsilon_y=1$ and $\Omega_z = 0$.
    The data represents
    a $\lb x>0,y=0,z=0 \rb$ cut with every point shown.
    The excited $n=1$ self-similar solution is shown~(solid line)
    with $\tau \equiv \ln | T^*- T|$ where $T^*$ is the time of collapse
    (so that collapse occurs at $\tau \rightarrow - \infty$).
    The collapse time of the $n=1$ solution is chosen so that the
    two solutions coincide for the first frame only.
    That they coincide for the other frames indicates the approach to
    the self-similar solution.
}
\end{figure}
%%%%%%%%%%%%%%%%% Figure:

\begin{figure}
\vspace{-5mm}
\scalebox{0.65}{\includegraphics{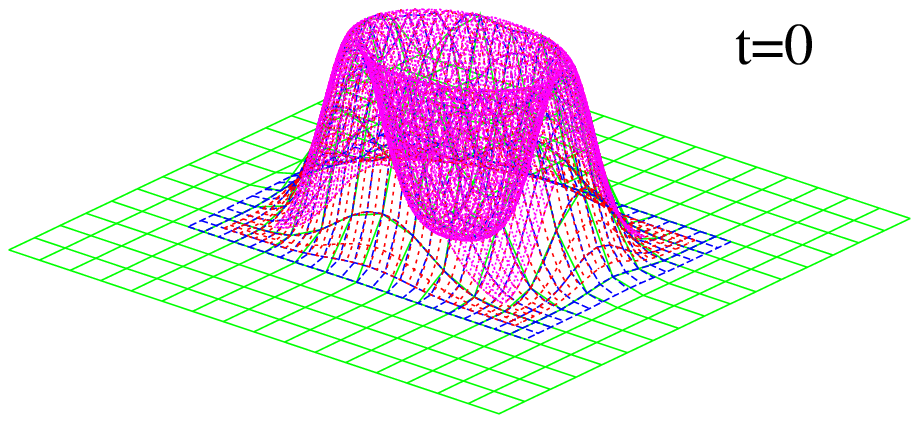}}\\
\vspace{-28mm}
\scalebox{0.65}{\includegraphics{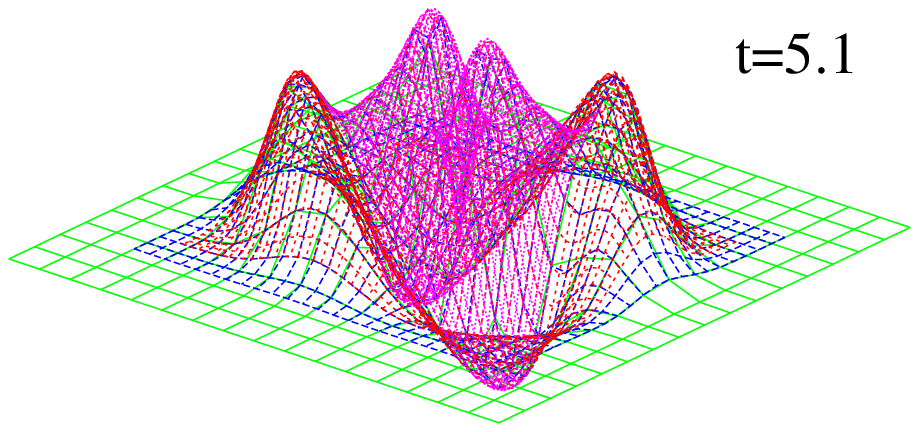}}\\
\vspace{-29mm}
\scalebox{0.65}{\includegraphics{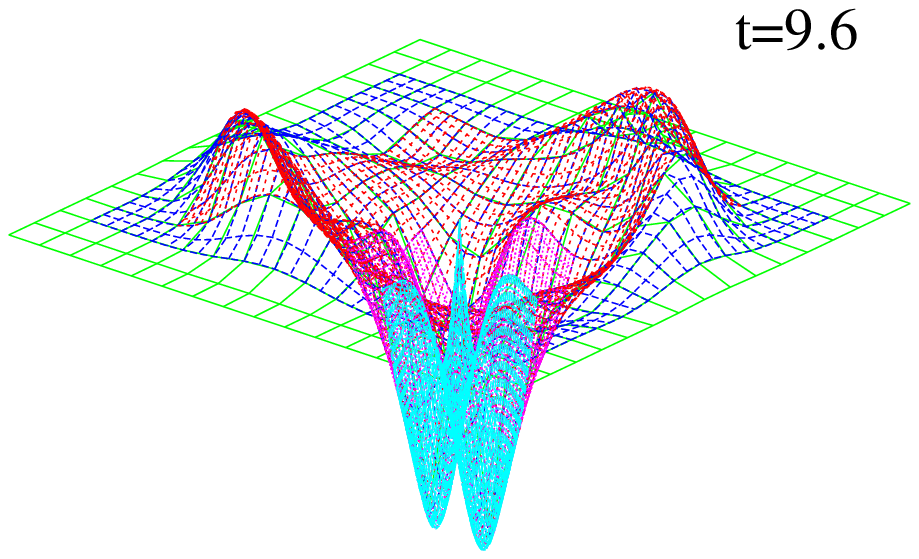}}\\
\vspace{-22mm}
\scalebox{0.65}{\includegraphics{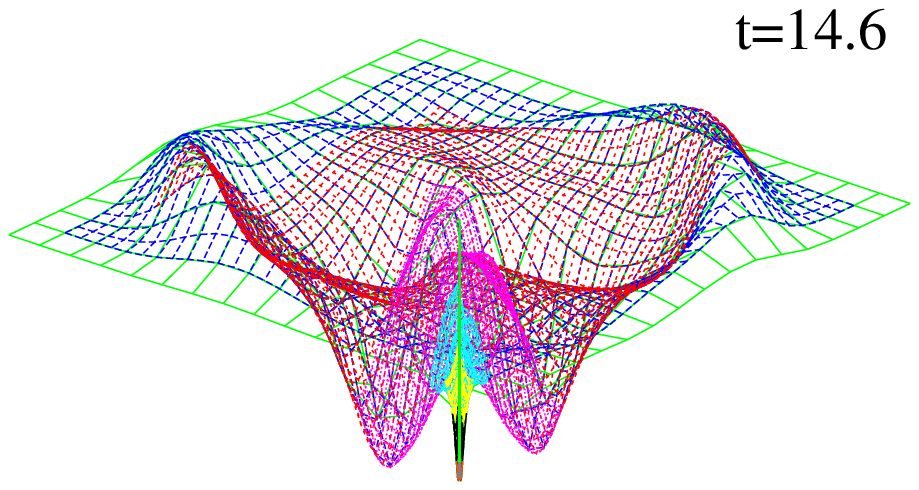}}
\vspace{-17mm}
\caption{\label{fig:omega} Snapshots of slightly sub-critical
   evolution. The field $\chi\lb t,x,y,z=0 \rb$ is shown at four times
   for a family of initial data
   with angular momentum about
   the $z$-axis. For clarity, only every other point in the
   $x$ and $y$ directions
   is shown
   (every fourth point).
   At the final time, all ten grids are shown
   with a refinement factor of 2.
   Greater detail of this last frame is shown in
   Fig.~\ref{fig:omega_zoom}.
   The initial data is a member of family~(a) from Table~\ref{table:init}
   with parameters
   $A\approx 1.3359$, $R=8$, $\delta=3$, $x_c=y_c=z_c=0$,
   $\epsilon_x = 0.5$, $\epsilon_y=1$, $\nu=0$, and $\Omega_z=0.4$.
   The dimensionless ratio of the angular momentum to the energy squared is $J/E^2 = 0.0025$.
}
\end{figure}

The usefulness of AMR depends obviously on whether it produces
correct solutions and whether it allows for
high resolution with only proportional work. Starting with the
unigrid code, convergence and energy conservation were confirmed.
The AMR results were then checked against high resolution
unigrid results.
Perhaps a stronger test is that this code finds the same results
for spherically symmetric initial data as that in~\cite{Liebling:1999nn}
(as demonstrated in Fig.~\ref{fig:ss}).
As an example of the benefit of AMR consider Fig.~\ref{fig:omega}. A rough
estimate of the computational work for the entire evolution is
$2^8$, in units where 
the work for a unigrid evolution at the coarsest resolution is unity.
In comparison, to achieve
uniform resolution equal to that of the finest sub-grid
would require work equal to $\left( 2^9 \right)^{3+1}$, a factor
$2^{26}$ more work than that with AMR.

%%%%%%%%%%%%%%%%% Figure:
%%%%%%%%%%%%%%%%% Figure:

%%%%%%%%%%%%%%%%% Figure:
\begin{figure}
\vspace{-3mm}
\scalebox{0.5}{\includegraphics{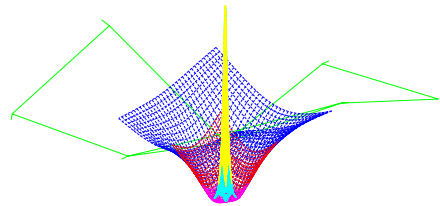}}
\vspace{-13mm}
\caption{\label{fig:omega_zoom} Detail of the peak region from
   the final frame of Fig.~\ref{fig:omega}. Only the finest four
   grids are shown (every point is shown) along with the coarsest grid
   in order to  compare the resolution achieved with respect to the coarse grid.
   The finest grid has a resolution  a factor of $2^9$ finer than the
   coarse grid.
}
\end{figure}
%%%%%%%%%%%%%%%%% Figure:

%%%%%%%%%%%%%%%%%%%%%%%%%%%%%%%%%%%%%%%%%%%%%%%%%%%%%%%%%%%%%%%%
{\bf \em Results:}
%%%%%%%%%%%%%%%%%%%%%%%%%%%%%%%%%%%%%%%%%%%%%%%%%%%%%%%%%%%%%%%%
As discussed in~\cite{Liebling:1999nn}, solutions generically evolve
to one of two stable end states. For large initial data, the energy density
collapses toward the origin suggesting the formation of a singularity.
As in the strictly spherical case, the usual caveat applies
that we only have indications, not proof, that a singularity is forming.
For small initial data, the energy density implodes through the
origin and then disperses to infinity.
Removing the restriction to spherical symmetry has
not revealed any additional stable end states.

The aim is to investigate the region of solution space separating
these two end states. In the jargon of nonlinear dynamics,
we look at the so-called {\em critical surface} occurring between
the two {\em basins of attraction}.
To do so, one chooses a 1-parameter
family of initial data with the property that data with small
parameter disperses while large parameter initial data forms
a singularity. By tuning the free parameter according
to the resulting end state of the evolution, the critical surface is approached.

The previous results in spherical symmetry indicated that all
initial data of compact support, when tuned,
approached the same universal, self-similar solution~\cite{Liebling:1999nn}.
This critical solution happens to be one of a family of solutions
found by Aminneborg and Bergstrom by assuming self-similarity
and solving the resulting ODE~\cite{Aminneborg:1995ff}. A linear perturbation
analysis revealed the $n=1$ solution to have the requisite single unstable
mode necessary to be a proper critical solution (a so-called {\em intermediate
attractor}).

This solution is shown in coordinates adapted to its self-similarity in
Fig.~\ref{fig:ss}. Given a self-similar solution which takes the form
$f(r,t) = f\lb r/| T^*-T|\rb$, one can recast it as
$g\lb\ln r - \ln|T^*-T| \rb$.
This form makes apparent that the solution executes linear motion
in log space and log time, where one is free to set the collapse time $T^*$.
Thus, in Fig.~\ref{fig:ss} the self-similar solution travels leftward
retaining its shape.

Along with the $n=1$ solution is shown a cut along the $x$-axis of a
near-critical evolution from
spherically symmetric initial data. Such an evolution would be expected
to agree with previous results in spherical symmetry, as it appears to
do in the plot. The collapse time of the $n=1$ solution
was chosen so that the two
solutions agree in the first frame. The $n=1$ solution is then completely
determined at other times by translation in $\ln r$.

It was also observed in~\cite{Liebling:1999nn} that whenever the range
of $\chi$
exceeded $\pi$ at a given time, a singularity would eventually form in its
future. However, evolutions of non-spherically symmetric families
do not obey such
a condition, having exceeded $\pi$ while remaining nonsingular.
One still finds solutions which disperse ({\em sub-critical}) 
separated from those which form a singularity ({\em super-critical}).

For such non-symmetric initial data, the question is whether some
different critical solution appears.
If the $n=1$ solution has non-symmetric unstable modes
not present in spherical symmetry, then
a change in critical solution would be expected. However, all such
non-symmetric variations of the types of initial data
in Table~\ref{table:init} have failed to find a different
critical solution. 

An example from initial data
with angular momentum is shown in Fig.~\ref{fig:omega}. The
pulse evolves in a nontrivial manner eventually dispersing much
of its energy toward the edges of the grid.  However, near the
origin, a spherically symmetric waveform emerges which propagates
toward the origin. In this case, the waveform has negative amplitude
(the model is invariant with respect to the transformation $\chi \rightarrow
-\chi$).
Further detail of this waveform is shown in Fig.~\ref{fig:omega_zoom}.
The upward sweep at the origin corresponds to the regularity constraint
mentioned previously that $\chi(0,0,0,t)=0$. 

%%%%%%%%%%%%%%%%% Figure:
\begin{figure}[h]
\vspace{-3mm}
\centerline{ \scalebox{0.45}{ \includegraphics{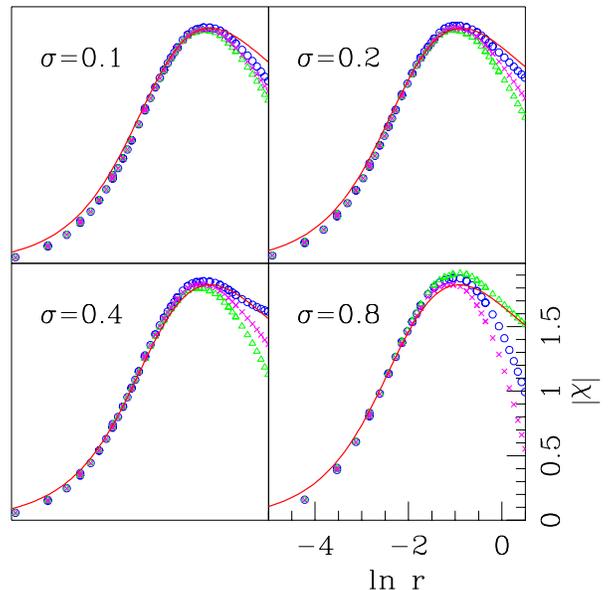} } }
\vspace{-1mm}
\caption{\label{fig:spherical} Demonstration of approach to a spherically
  symmetric solution at criticality.
  Initially asymmetric initial data is tuned and slightly sub-critical
  solutions are shown above for different families.
  Here $\sigma$ is a measure of
  the asymmetry ($\sigma = 0$ corresponds to spherical symmetry)
  which partially defines family (a) in Table~\ref{table:init}
  via $\epsilon_x = 1 + \sigma$ and
  $\epsilon_y = 1 - \sigma$.
  For each value of $\sigma$, three slices are shown:
  $(x>0,0,0)$ (circles), $(0,y>0,0)$ (triangles),
  and $(0,0,z>0)$ (crosses). Their general agreement near the origin
  demonstrates approach to spherical symmetry. Also shown 
  (line) is
  the $n=1$ self-similar solution (with collapse time adjusted to match
  the solution).
}
\end{figure}
%%%%%%%%%%%%%%%%% Figure:

%%%%%%%%%%%%%%%%% Figure:
\begin{figure}[ht]
\vspace{-7mm}
\scalebox{0.65}{\includegraphics{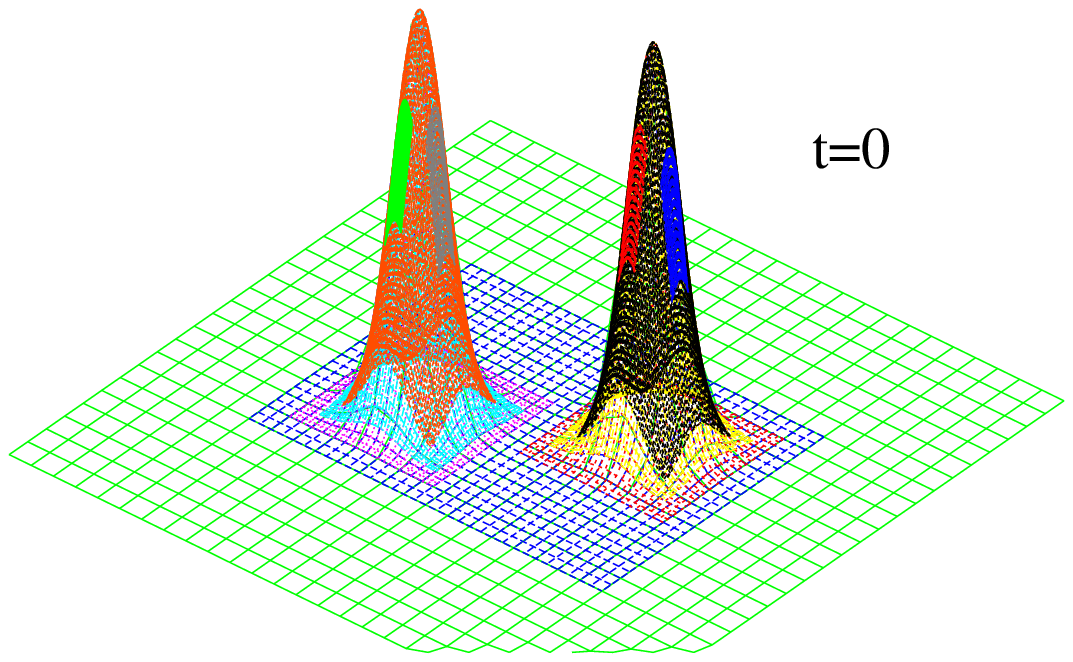}}\\
\vspace{-21mm}
\scalebox{0.65}{\includegraphics{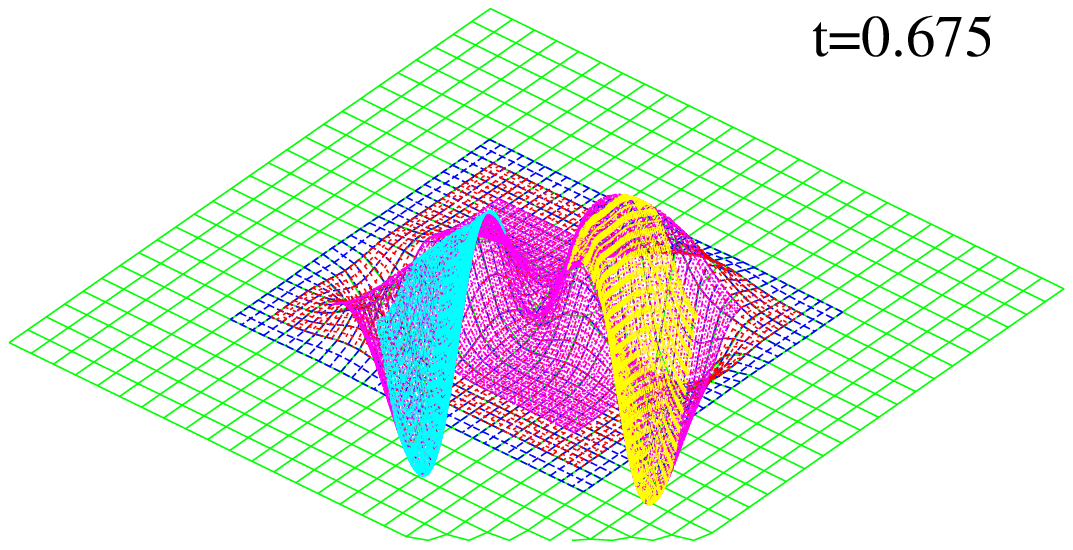}}\\
\vspace{-21mm}
\scalebox{0.65}{\includegraphics{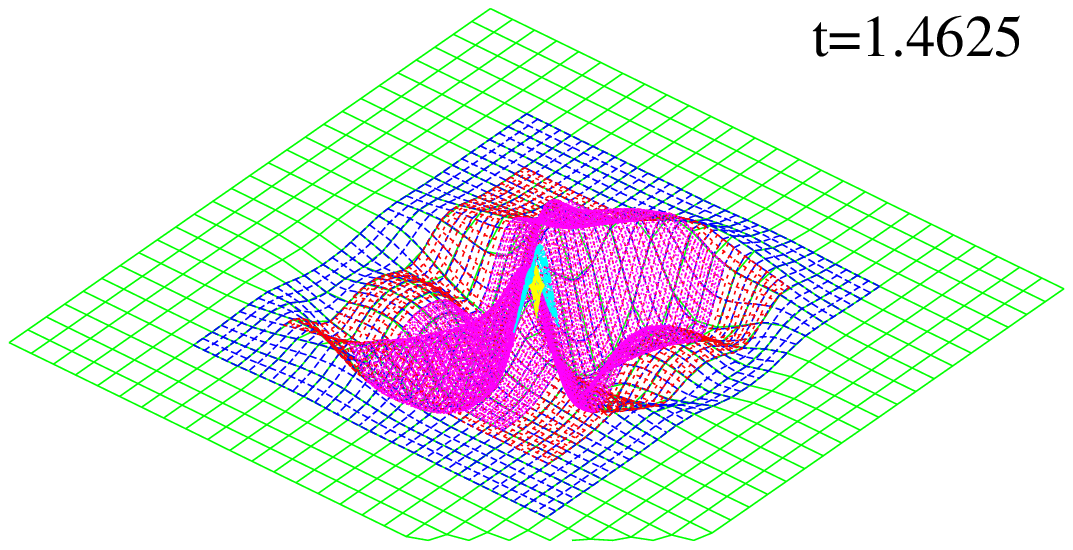}}\\
\vspace{-21mm}
\scalebox{0.65}{\includegraphics{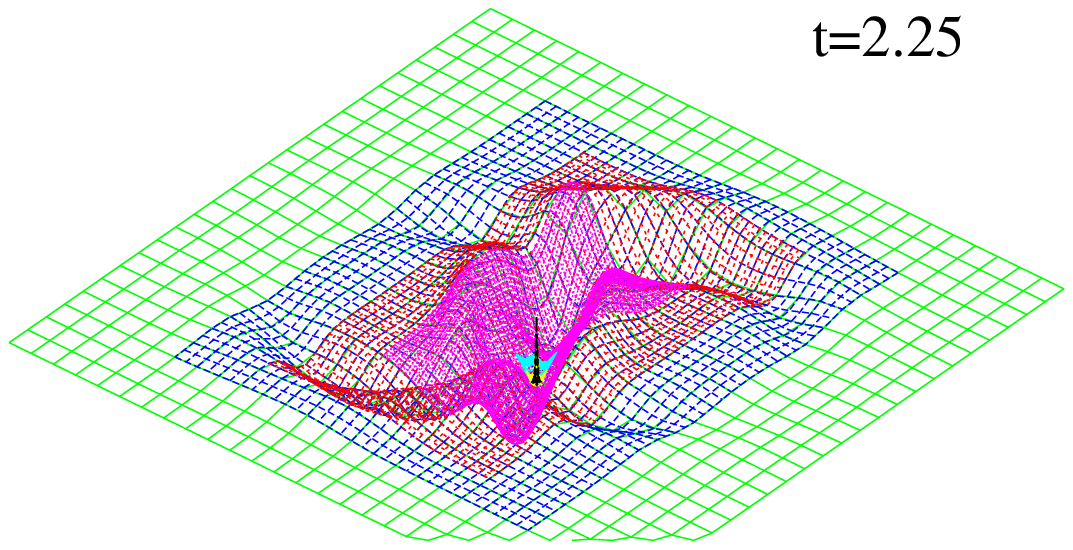}}
\vspace{-9mm}
\caption{\label{fig:twopulses} Snapshots of a slightly sub-critical
   collision of two grazing pulses.
   The field $\chi\lb t,x,y,z=0 \rb$ is shown in the $x-y$ plane
   at four times for two pulses with linear momentum in $\pm x$ directions.
   At late times, the region near the origin (the fine grids in the center
   of the final two frames) rapidly approaches
   spherical symmetry and self-similarity.
   The initial data consists of family~(b) from Table~\ref{table:init}
   with parameters $A_1 = A_2 = 11.5$, $R_1=R_2=0$, $\delta_1=\delta_2=0.5$,
   $\left(x_{c1},y_{c1},z_{c1}\right) = \left(-x_{c2},-y_{c2},z_{c2}\right)
    =\left(1.4,0.7,0\right)$, $\epsilon_{x1}=\epsilon_{x2}=1$, $\epsilon_{y1}
    = \epsilon_{y2} = 1$, and $v_1=-v_2=1$.
   The dimensionless ratio of the angular momentum to the energy squared is $J/E^2 = 0.0008$.
}
\end{figure}
%%%%%%%%%%%%%%%%% Figure:

Another example of initial data with angular momentum is
shown in Fig.~\ref{fig:twopulses}. Two pulses are sent towards one
another and, for sufficiently large initial amplitude, singularity
formation occurs. Four frames of a near critical evolution show
the region near the origin approaching the spherically symmetric,
self-similar solution.

A more systematic exploration away from spherical symmetry is shown
in Fig.~\ref{fig:spherical}. Here, $\sigma$ parameterizes the asymmetry
of the initial data, and near-critical solutions are shown for four
different values of $\sigma$. Furthermore, for each case the region
near the origin (the region which approaches self-similarity) is
shown along the three different axes. Their agreement suggests that
this region
approaches spherical symmetry.

There remains, however, the possibility that some other spherically
symmetric self-similar solution becomes the critical solution away
from spherical symmetry. Though there is no evidence of any transition,
the various excited self-similar solutions of~\cite{Aminneborg:1995ff}
are very similar near the origin consisting of a peak
followed by oscillations about $\chi=\pi/2$. To distinguish among
them, a perturbation analysis allowing for non-symmetric modes
would be appropriate, with the
critical solution being the one with
a single unstable mode.

The evidence therefore suggests that the $n=1$
self-similar solution remains the intermediate attractor away
from spherical symmetry, but a perturbation analysis
should settle the issue.
Further work will likely be directed towards
(1) using the AMR infrastructure with a gravitating model,
(2) relaxing the noted simplifications in the clustering,
and
(3) distributing the grids using MPI along the lines of~\cite{Hern:1999dq}.

%%%%%%%%%%%%%%%%%%%%%%%%%%%%%%%%%%%%%%%%%%%%%%%%%%%%%%%%%%%%%%%%
\begin{acknowledgments}
{\bf \em Acknowledgments:} 
%%%%%%%%%%%%%%%%%%%%%%%%%%%%%%%%%%%%%%%%%%%%%%%%%%%%%%%%%%%%%%%%
Many thanks to Frans Pretorius and Matthew~W. Choptuik for providing me
with the extremely useful Data-Vault, a software application they are developing
for visualizing
AMR data. Thanks go also to Richard~A. Matzner for
reviewing the manuscript.
I appreciate 
the support of NSF PHY-9900644 and
of the financial support of Southampton College.
\end{acknowledgments}

%%%%%%%%%%%%%%%%%%%%%%%%%%%%%%%%%%%%%%%%%%%%%%%%%%%%%%%%%%%%%%%%
\bibliography{paper}
%%%%%%%%%%%%%%%%%%%%%%%%%%%%%%%%%%%%%%%%%%%%%%%%%%%%%%%%%%%%%%%%

\end{document}